\pdfoutput=1
\documentclass[sigconf,compact]{acmart}

\makeatletter
\def\mdseries@tt{m}
\makeatother

\usepackage{
    graphicx,
	url,
    hyperref,
    cleveref,
    adjustbox,
    float,
    multirow,
    tabularx
}
\usepackage[frozencache,cachedir=.]{minted}
\usepackage[acronym,nopostdot,nogroupskip,nonumberlist]{glossaries}
\def\BibTeX{{\rm B\kern-.05em{\sc i\kern-.025em b}\kern-.08emT\kern-.1667em\lower.7ex\hbox{E}\kern-.125emX}}
    
\copyrightyear{2020}
\acmYear{2020}
\setcopyright{ccby}
\acmConference[CARRV 2020]{Fourth Workshop on Computer Architecture Research with RISC-V}{May 29, 2020}{Virtual Workshop}
\acmDOI{}
\acmISBN{}

\newcommand{\rvm}{R2VM}

\newacronym{ISS}{ISS}{instruction set simulator}
\newacronym{RTL}{RTL}{register-transfer level}
\newacronym{TLB}{TLB}{translation-lookaside buffer}
\newacronym{ISA}{ISA}{instruction set architecture}
\newacronym{PC}{PC}{program counter}
\newacronym{IPC}{IPC}{instruction per cycle}
\newacronym{DBT}{DBT}{dynamic binary translation}
\newacronym{LRU}{LRU}{least-recently used}
\newacronym{IPI}{IPI}{inter-processor interrupt}
\newacronym{SBI}{SBI}{supervisor binary interface}
\newacronym{OS}{OS}{operating system}
\newacronym{CSR}{CSR}{control and status register}

\begin{document}

\title[Accelerate Cycle-Level Multi-Core RISC-V Simulation with Binary Translation]{Accelerate Cycle-Level Full-System Simulation of Multi-Core RISC-V Systems with Binary Translation}

\author{Xuan Guo}
\email{Gary.Guo@cl.cam.ac.uk}
\affiliation{%
  \institution{University of Cambridge}
  \city{Cambridge}
  \country{UK}
}

\author{Robert Mullins}
\email{Robert.Mullins@cl.cam.ac.uk}
\affiliation{%
  \institution{University of Cambridge}
  \city{Cambridge}
  \country{UK}
}

\renewcommand{\shortauthors}{Guo, et al.}

%
\begin{abstract}
It has always been challenging to balance the accuracy and performance of \glspl{ISS}. \Gls{RTL} simulators or systems such as gem5 \cite{binkert2011gem5} are used to execute programs in a cycle-accurate manner but are often prohibitively slow. In contrast, functional simulators such as QEMU \cite{bellard2005qemu} can run large benchmarks to completion in a reasonable time yet capture few performance metrics and fail to model complex interactions between multiple cores. This paper presents a novel multi-purpose simulator that exploits binary translation to offer fast cycle-level full-system simulations.
Its functional simulation mode outperforms QEMU and, if desired, it is possible to switch between functional and timing modes at run-time. 
Cycle-level simulations of RISC-V multi-core processors are possible at more than 20 MIPS, a useful middle ground in terms of accuracy and performance with simulation speeds nearly 100 times those of more detailed cycle-accurate models.

\end{abstract}

%
%

%

\maketitle

\section{Introduction}

RISC-V is a free, open, and extensible ISA. With the ongoing ecosystem development of RISC-V and an increasing number of companies and institutions switching to RISC-V for both production and research, RISC-V has become the test bed instruction set of computer architecture research. A key tool when exploring new architectural trade-offs is the instruction-set simulator (ISS). Fast cycle-level simulation allows new ideas to be validated quickly at an appropriate level of abstraction and without the complexities of hardware development. In particular we focus on the challenge of simulating multi-core RISC-V systems.

The design of a processor can broadly be divided into the design of the core and memory subsystem. Characterising the performance of the core pipeline in isolation is often a simpler task to that of characterising the memory system. While smaller synthetic benchmarks are useful at the core level, larger more complex and longer running workloads are often needed to understand the memory system and the potential interactions between cores.

For example, the smaller synthetic MCU benchmark CoreMark \cite{coremark} executes at a magnitude of $10^{8}$ instructions per iteration, while SPEC2017~\cite{spec17}, a larger and more realistic benchmark running real-life applications, requires a magnitude of $10^{12}$ instructions for a single run \cite{limaye2018workload}. SPEC takes from hours to days to run even on real machines, and hardly possible for simulators to run to completion. It is therefore helpful to have a fast simulator.

Of course, fast simulation involves a trade-off between the fidelity of the model and the speed at which simulations can be completed. Unfortunately, we are currently forced to choose between slow cycle-accurate simulators or fast functional-only simulators such as QEMU. In particular, there is a lack of fast full-system simulators that can accurately model cache-coherent multi-core processors.

In this paper, we present the Rust RISC-V Virtual Machine (\rvm{}). \rvm{} is written in the increasingly popular high-level system programming language Rust \cite{rust}. \rvm{} is released \footnote{Available at \url{https://github.com/nbdd0121/r2vm}.} under permissive MIT/Apache-2.0 licenses in the hope to encourage its adoption and expansion by the broader community. To our knowledge, this is the first binary translated simulator that supports cycle-level multi-core simulation. It can accurately model cache coherency protocols and shared caches. Cycle-level simulations are possible at more than 20 MIPS, while the performance of functional-only simulations can outperform QEMU and exceed 400 MIPS.  

\section{Background}
\label{sec:related}

\subsection{Instruction Set Simulators}

\Glspl{ISS} can be classified as either execution-driven, emulation-driven or trace-driven \cite{brais2020survey}. We omit a detailed discussion of execution-driven
simulators such as Cachegrind \cite{nethercote2007valgrind} that modify programs with binary instrumentation and execute them natively, because they require the host and the guest ISA to be identical and do not support full-system simulation. Emulation-driven simulators emulate the program execution, and gather performance metrics on-the-fly; in contrast, trace-driven simulators run emulation before-hand and gather traces from the program, e.g. branches or memory accesses, and later replay the trace against a specific model. Traces allow ideas to be evaluated quickly without the need to simulate in detail, but cannot easily capture effects that may alter the instructions that are executed, e.g. inter-core interactions or speculative execution \cite{brais2020survey}. Moreover, storage space required for traces grows linearly with the length of execution, making trace-driven simulators incapable of simulating large benchmarks. \rvm{} is an emulation-driven simulator and the remainder of this paper will focus exclusively on emulation-driven simulators.

Simulators can also be categorised by their levels of abstraction. One category of simulators is functional simulators. Functional simulators simulate the effects of instructions without taking microarchitectural details into account. Because less information is needed, aggressive optimisations can be performed, and the performance is usually several magnitudes faster than timing simulators. QEMU falls into this category. It should be noted though that while QEMU itself is a purely functional simulator, it can be modified to collect metadata for off-line or on-line cache simulation \cite{van2014cache}.

The other category is timing simulators. Out of timing simulators, \gls{RTL} simulators can model processor microarchitectures very precisely, but the difficulty in implementing a feature in \gls{RTL} simulator is not much different from implementing it in hardware directly. \gls{RTL} simulators are also poor in performance, usually run at a magnitude of kIPS \cite{ta2018simulating}.

At a higher level, there are cycle-level microarchitectural simulators. These
are able to omit \gls{RTL} implementation details to improve performance while retaining a detailed microarchitectural model. An popular example is the gem5 simulator running with In-Order or O3 mode \cite{binkert2011gem5}. For faster performance, we can give up some extra microarchitectural details and predict the number of cycles taken for each non-memory instruction instead of computing them in real-time, and in the extreme case, assume all non-memory operation only takes 1 cycle to execute as gem5's ``timing simple'' CPU model assumes. This approach is no longer cycle-accurate, but this cycle-approximate model is often adequate to perform cache and memory simulations.

\subsection{Binary Translation}

Binary translation is a technique that accelerates \gls{ISA} simulation or program instrumentation \cite{hazelwood2011dynamic}. An interpreter will fetch, decode and execute the instruction pointed by the current \gls{PC} one-by-one, while binary translation will, either ahead of time (static binary translation) or in the runtime, i.e. when the block of code if first executed (\gls{DBT}), translate one or more basic blocks from the simulated \gls{ISA} to the host's native code, cache the result, and use the translation result next time the same block is executed.

QEMU uses binary translation for cross-\gls{ISA} simulation or when there is no hardware virtualisation support \cite{bellard2005qemu}. Böhm et al. proposed a method to introduce binary translation to single-core timing simulation in 2010 \cite{bohm2010cycle}.

\subsection{Multi-core Simulation}

Extending single core simulators to handle multiple cores is complicated by the performance implications of the ways in which cores may interact. As cores share caches and memory, simulations of individual cores cannot simply be run independently. For example, accurate modelling of cache coherence, atomic memory operations and \glspl{IPI} must be considered.

Böhm et al.'s modified ARCSim simulator \cite{bohm2010cycle} can model single-core processors with high accuracy and reasonable performance; however, Almer et al.'s extension to Böhm et al.'s work \cite{almer2011scalable} that essentially runs multiple copies of the single-core simulator in parallel threads to provide multi-core support is limited in its fidelity. The author comments ``detailed behaviour of the shared second-level cache, processor interconnect and external memory of the simulated multi-core platform'' cannot be modelled accurately. QEMU is able to exploit multiple cores to emulate a multi-core guest but provides only a functional simulation mode and supports no timing or modelling of the memory system.

An accurate model of cache coherence and the memory hierarchy requires that multiple cores are simulated in lockstep (or in a way that guarantees equivalent results). 
Simulators that forego this are unable to properly simulate race conditions and shared resources. Existing cycle-level simulators such as gem5 achieve lockstep by iterating through all simulated cores each cycle. This causes a significant performance drop. Spike (or \texttt{riscv-isa-sim}), on the other hand, switches the active core to simulate less frequently. Its default compilation option only switches the core every 1000 cycles, making it impossible to model race conditions where all cores are trying to acquire a lock simultaneously. No existing binary translated simulators can model multi-core interaction in lockstep, and therefore none of these can model cache coherency or shared second-level cache properly.

\section{Implementation}
\label{sec:impl}

\subsection{Overview}
\label{sec:overview}

\begin{figure*}[ht]
    \includegraphics[scale=0.68]{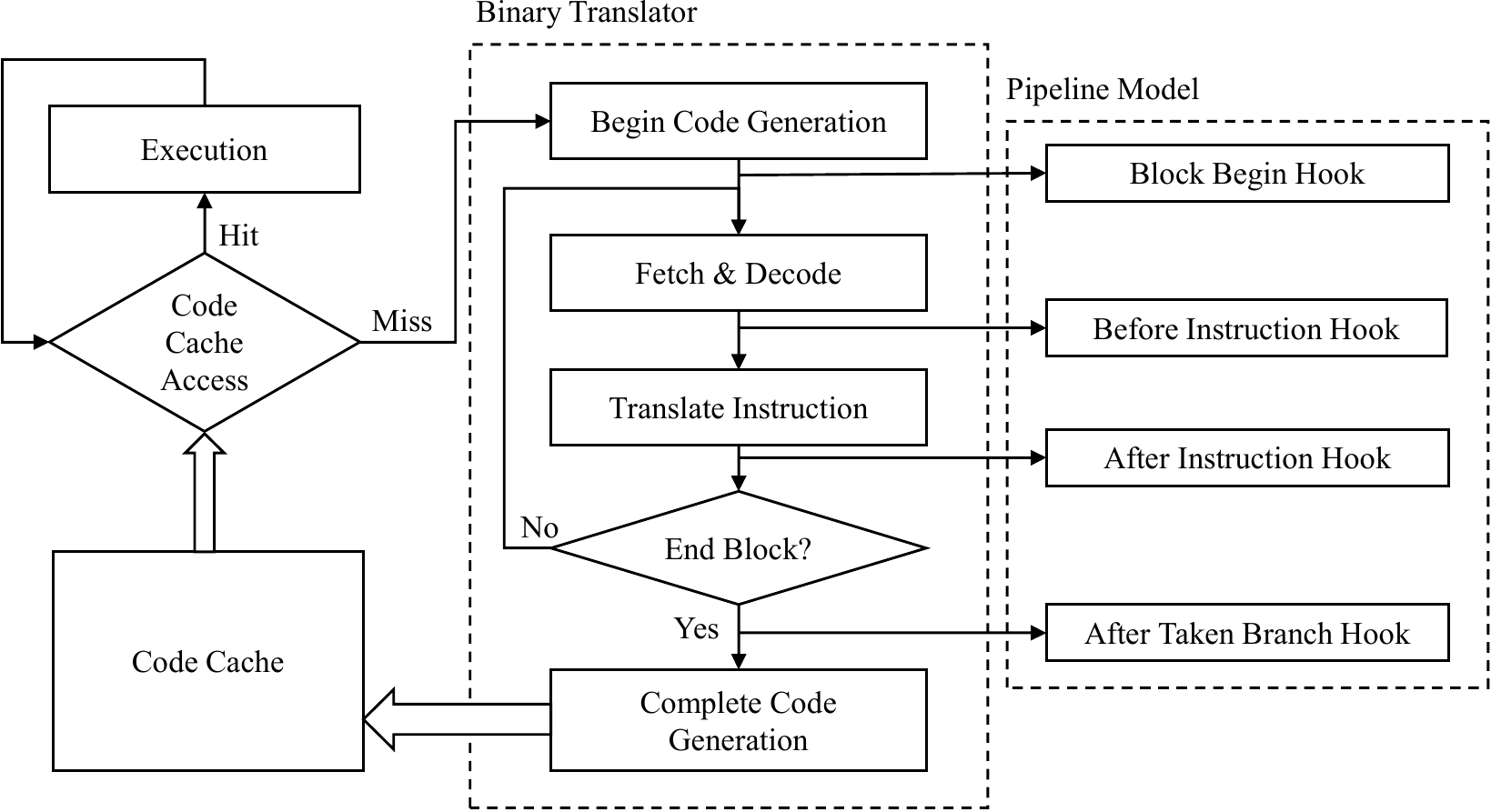}
    \caption{Control flow overview of the simulator}
    \label[figure]{fig:control}
\end{figure*}

The high-level control flow of \rvm{}, as shown in \Cref{fig:control}, is similar to other binary translators. When an instruction at a particular \gls{PC} is to be executed, the code cache is looked up and the cached translated binary is executed directly if found; otherwise, the binary translator is invoked and an entire basic block is fetched, decoded, and translated. We have used a variety of techniques to improve the binary translator performance that are often found in other binary translators, such as block chaining~\cite{bellard2005qemu}.

As full-system simulation is supported, we have to deal with the case that a 4-byte uncompressed instruction spans two pages. We handle this by creating a stub that reads the 2 bytes that lie on the second page each time the stub is executed, and patches the generated code if 2 bytes read are different from that of initial translation.

Cota et al.~\cite{cota2019cross} suggests sharing a code cache between multiple cores to promote code reuse and boost performance. In contrast, we provide each hardware thread its own code cache. This allows different code to be generated for each core, e.g. in the case of heterogeneous cores. This also lessens the synchronisation requirements when modifying the code cache, simplify the implementation.

\subsection{Pipeline Simulation}
\label{sec:pipeline}

The main difference between our simulator's flow and existing ones such as QEMU is that we introduce ``pipeline model''s, which comprises several hooks. Hooks can process relevant instructions and generate essential microarchitectural simulation code if necessary. The hooks can also indicate the number of cycles it would take for the instruction to complete. It should be noted that this is only for the execution pipeline, while memory systems and cache are in a separate component.

For simple models, such as gem5's ``timing simple'' model where each instruction takes 1 cycle to execute, implementation is straightforward as shown in \Cref{listing:simple}.

\begin{listing}[ht]
\begin{minted}[linenos,breaklines]{rust}
#[derive(Default)]
pub struct SimpleModel;

impl PipelineModel for SimpleModel {
    fn after_instruction(&mut self, compiler: &mut DbtCompiler, _op: &Op, _compressed: bool) {
        compiler.insert_cycle_count(1);
    }

    fn after_taken_branch(&mut self, compiler: &mut DbtCompiler, _op: &Op, _compressed: bool) {
        compiler.insert_cycle_count(1);
    }
}
\end{minted}
\caption{Timing simple model implementation}
\label[listing]{listing:simple}
\end{listing}

We have also implemented and validated an in-order pipeline model that accurately models a classic 5-stage pipeline with a static branch predictor. Our implementation captures pipeline hazards, such as data hazards caused by load-use dependency and stalls due to a branch/jump into a misaligned 4-byte instruction. Unlike Böhm et al.'s simulator \cite{bohm2010cycle} which needs to call a ``pipeline'' function after each instruction, our implementation models pipeline behaviours during \gls{DBT} code generation and reflects them as number of cycles taken, therefore requires no explicit code to be executed in runtime.

More complex processors may need either to make an estimation of pipeline states (and sacrifice some accuracy) or generate custom assembly in the hooks to maintain these states during execution (and sacrifice some performance).

\subsection{Multi-core Simulation}

The techniques we described in the previous section works well for single-core systems. But as described in the background section, running them in parallel or switch between them in a coarse-grained manner has a huge impact on simulation accuracy of multi-threaded programs. The ideal scheduling granularity is, therefore, a cycle, i.e. having all simulated cores run in lockstep. This is, however, difficult to achieve for binary translators.

We experimented the idea of using thread barriers to synchronise multiple threads each simulating a single core. It turns out we could only synchronise 1 million times per second even after careful optimisation at the assembly level.

\subsubsection{Lockstep Simulation}

The approach we use takes inspiration from fibers, sometimes also referred to as coroutines or green threads. Fibers are cooperatively scheduled by the user-space application, and they voluntarily ``yield'' to other fibers, in contrast to traditional threads which are preemptively scheduled by the operating system and are generally heavy-weight constructs. Fibers are often used in I/O heavy, highly concurrent workloads, such as network programming, but this time we borrowed it to our simulator.

In our implementation, we create one fiber for each hardware thread simulated, plus a fiber for the event loop. Each time the pipeline model instructs the \gls{DBT} to wait for a few cycles, we will generate a number of yields. \Cref{listing:example} shows an example of generated code under timing simple model.

\begin{listing}[ht]
\begin{minted}[linenos,breaklines]{nasm}
mov     rax, qword [rbp+0x78] ; \
mov     qword [rbp+0x70], rax ; | add a4, zero, a5
call    fiber_yield_raw       ; /
mov     eax, dword [rbp+0x78] ; \
add     eax, -0x1             ; |
cdqe                          ; | addiw a5, a5, -1
mov     qword [rbp+0x78], rax ; |
call    fiber_yield_raw       ; /
mov     eax, dword [rbp+0x70] ; \
imul    eax, dword [rbp+0x50] ; |
cdqe                          ; | mulw a0, a4, a0
mov     qword [rbp+0x50], rax ; |
call    fiber_yield_raw       ; /
\end{minted}
\caption{Example of generated code with yield calls. RBP points to the array of RISC-V registers.}
\label[listing]{listing:example}
\end{listing}

\begin{figure*}[ht]
    \includegraphics[scale=0.7]{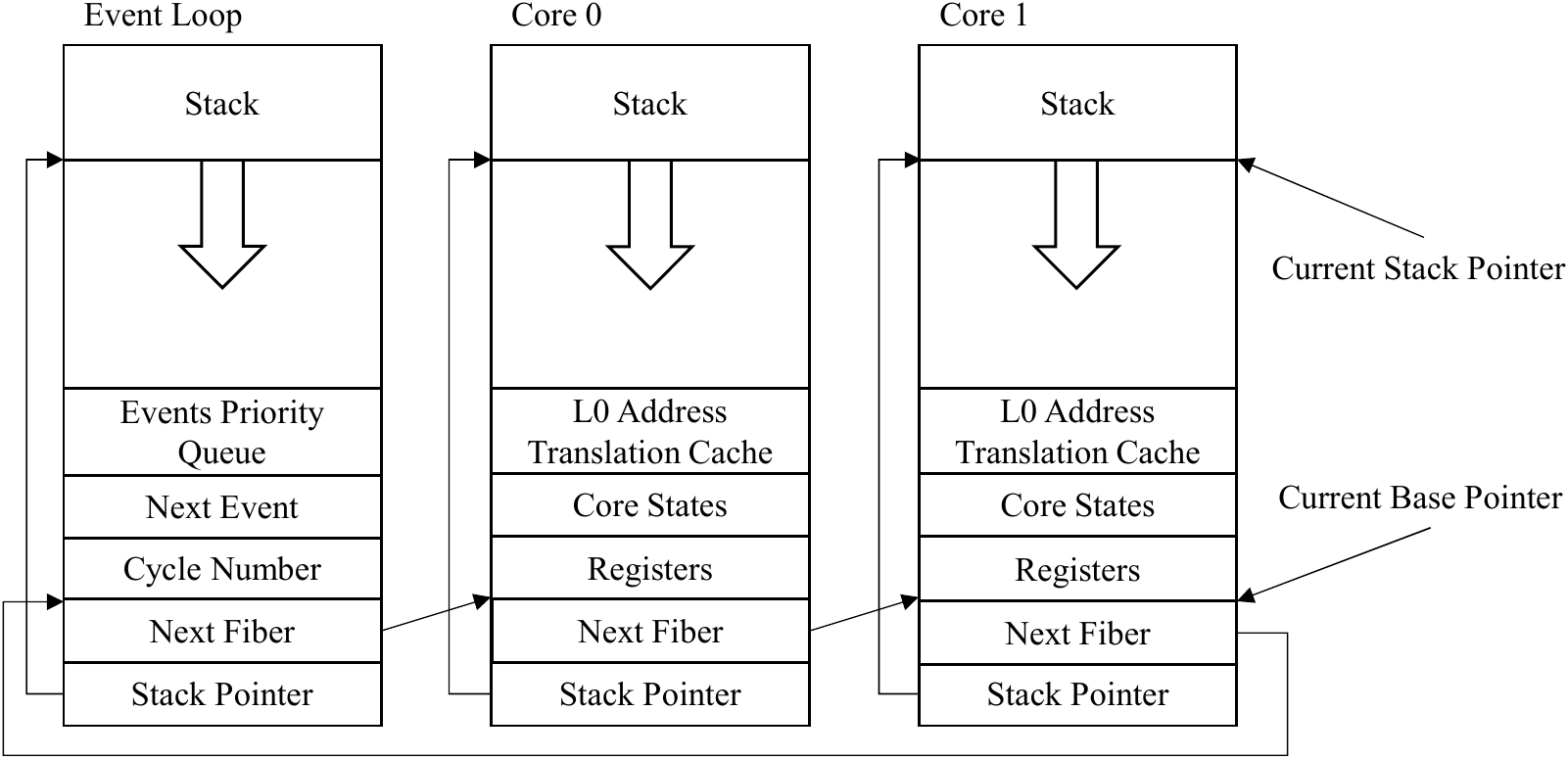}
    \caption{Memory layout of fibers}
    \label[figure]{fig:layout}
\end{figure*}

Different from normal fiber implementation, we engineered the fiber's memory layout to look like \Cref{fig:layout} to suit the need of a simulator. Each fiber is allocated with a 2M memory aligned to 2M boundary, and the stack for running code under the fiber is contained within this memory range. The alignment requirement allows the fiber's start address to be recovered from the stack pointer by simply masking out least significant 21 bits. The base pointer points to the end of fixed fiber structures rather than the beginning, so that positive offsets from the base pointer can be used freely by the \gls{DBT}-ed code, while the negative offsets are used for fiber management.

The ABI of the host platform for \gls{DBT}-ed code is not respected; we rather specify all registers other than the base pointer and stack pointer to be volatile, or caller-saved. By doing so, \texttt{fiber\_yield\_raw} does not need to bear the cost of saving any registers. To yield in non-\gls{DBT}-ed code, we can alternatively push ABI-specified callee-saved registers into the stack and switch.

This careful design makes fiber switching lightning fast; the \texttt{fiber\_yield\_raw} function is as simple as 4 instructions on AMD64, shown in \Cref{listing:yield}.

\begin{listing}[ht]
\begin{minted}[linenos,breaklines]{nasm}
fiber_yield_raw:
    mov [rbp - 32], rsp ; Save current stack pointer
    mov rbp, [rbp - 16] ; Move to next fiber
    mov rsp, [rbp - 32] ; Restore stack pointer
    ret
\end{minted}
\caption{Implementation of the fiber yielding code}
\label[listing]{listing:yield}
\end{listing}

\subsubsection{Synchronisation Points}

Simply yielding a few cycles after every executed instruction will severely limit performance and in many cases will be unnecessary. In practice, we only need to synchronise at points where the execution pipeline can produce visible side-effects to other cores and/or the rest of the system, or where the rest of the system's behaviour would affect the running pipeline.

We observe that there are three ways that a pipeline interacts with another:
\begin{itemize}
    \item An memory operation is performed.
    \item An control register operation is performed. This includes read/write to performance monitor registers, or control registers related to the memory system.
    \item An interrupt happens.
\end{itemize}

For the first two types of interaction, we insert a synchronisation point before and after they are executed. For the third case (interrupts), because it is generally difficult to interrupt an \gls{DBT}-ed code mid-way, we choose to check for interrupts only at the end of basic blocks. We believe that this decision will not affect the accuracy of our simulation due to the inherent entropy of I/O operations.

The positioning of yielding that lies in between two synchronisation points, therefore, would have no visible side-effects and cannot be distinguished. Our implementation postpones all yielding until the next synchronisation point. We tweaked our yield implementation as shown in \Cref{listing:yield} slightly to allow multi-cycle yield, and it demonstrates around 10\% performance gain compared to naive yielding.

\subsection{Memory Simulation}

\begin{figure*}[ht]
    \includegraphics[scale=0.7]{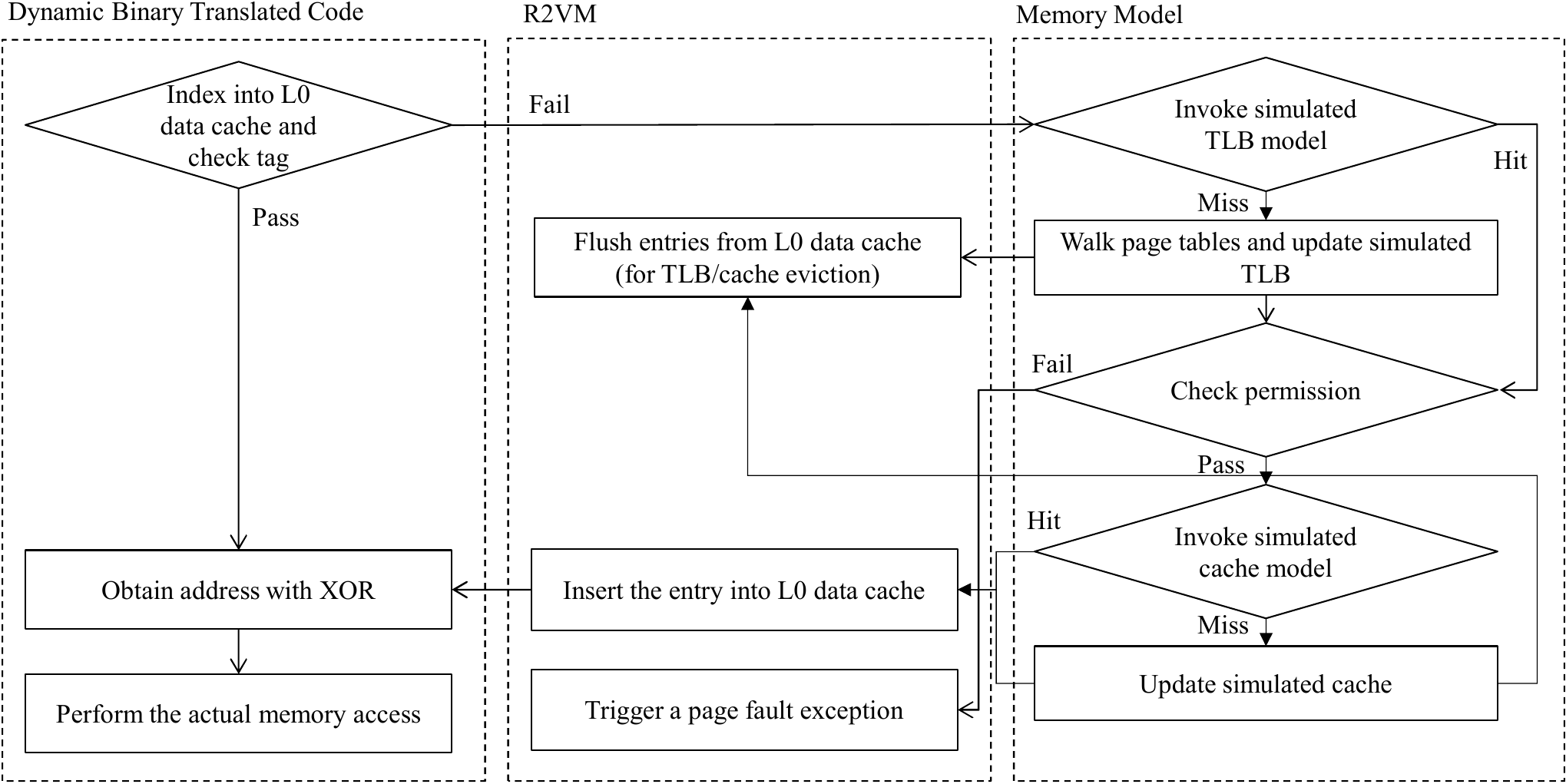}
    \caption{Control flow for memory accesses}
    \label[figure]{fig:memflow}
\end{figure*}

Previous sections described how we simulate each core's processing pipeline and how we achieve simulation in lockstep. The techniques we described and implemented speeds up pipeline simulation, but the speedup could be very limited when all memory accesses are still simulated. Moreover, the instruction cache and \glspl{TLB} would also need to be simulated for accuracy.

\subsubsection{L0 Data Cache}

For memory operations, each running core has its own ``L0 data cache''. When a core needs to read from or write to a memory address, it first checks if it is in the L0 data cache. If it hits, then memory access is performed entirely within \gls{DBT}-ed code, bypassing the memory model entirely.

As a result, the memory model will not intercept all memory accesses. It is therefore important to control what could be in the L0 data cache. We maintain a property that if an access hits the L0 data cache, then it must be a cache hit would the memory access reach the memory model. We speed up \gls{TLB} simulation with a similar approach in our previous work \cite{guo2019fast}.

In our previous \gls{TLB} simulation work, the property mandates an invariant that all entries in L0 \gls{TLB} are in the L1 data \gls{TLB}. The invariant kept in \rvm{} is that all L0 data cache entries are contained both in L1 data \gls{TLB} and L1 data cache. Therefore, as shown in \Cref{fig:memflow}, when entries are evicted from either the simulated \gls{TLB} or cache model, corresponding entries need to be flushed from the L0 data cache for the inclusiveness property.

\begin{figure}[ht]
    \includegraphics[scale=0.68]{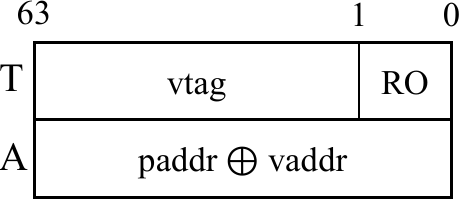}
    \caption{Memory layout of a tag entry in L0 data cache}
    \label[figure]{fig:l0d}
\end{figure}

We carefully engineered the memory layout of L0 data cache entries for maximum efficiency. The L0 data cache is direct-mapped, with each entry representing a cache line. Each entry has a memory layout like \Cref{fig:l0d}. It does not store actual memory contents; it rather stores a translation from the virtual tag to a physical address. In a sense, it is more like a \gls{TLB} with cache-line granularity than a cache. We pack the XOR-ed value of guest physical address and corresponding guest virtual address, plus a bit indicating if the cache line is read-only to a single machine word.

For each memory access, the L0 data cache is indexed into using the virtual tag. For read access, we compare if \verb|T >> 1| is equal to vtag. For write access, we compare if \verb|vtag << 1| is equal to \verb|T|. If the check passes, the requested virtual address is XOR-ed with \verb|A| to produce the address to access directly within \gls{DBT}-ed code. If the check fails, the cold path is executed and the memory model is invoked. The memory model will simulate both \gls{TLB} and data cache, and either triggers a page fault or inserts an entry into the L0 data cache.

The existence of L0 data cache promises the performance of \rvm{}'s fast-path, because it requires only 3 memory operations for each memory operation simulated. In the default configuration, because the memory model does not intercept all memory accesses, replacement policies such as \gls{LRU} cannot be used for the simulated \gls{TLB} and cache. Generally, we believe this is an acceptable accuracy loss to trade vastly better simulation performance. If \gls{LRU}-like policies are really needed, the L0 data cache could be bypassed and the memory model be invoked for each memory access, in sacrifice of performance.

\subsubsection{L0 Instruction Cache}
\label{sec:l0i}

\rvm{} also simulates instruction \gls{TLB} and caches similar to the data cache. Each core also has its own L0 instruction cache, with a simpler entry layout because read/write permission needs not to be distinguished. To keep the overhead of simulating instruction cache down, instead of accessing it each time an instruction is executed, we instead do it only when a basic block begins, or when the instruction being translated is in a different cache line compared to the previous instruction. For a cache line size of 64 bytes, this means that we only need to generate a single L0 instruction cache access for every 16-32 instructions.

We also creatively use the L0 instruction cache to optimise jumps across pages. Traditionally, because the page mapping might change and therefore the actual target of jump instruction might change, \gls{DBT}s have to conservatively not link these blocks together. We instead check the L0 instruction cache (which we would need to check anyway when next block begins) and see if the target is the same as the cached target. If so, the cached target is used and the control does not go back to the main loop.

\subsubsection{Cache Coherency}

Our design for the memory system inherently supports the use of cache coherency. Whenever the cache coherency protocol requires an invalidation, it can be flushed from the L0 data cache of the target core. Because all simulated cores execute in lockstep, and there are synchronisation points before all memory accesses, the effect of the invalidation will be visible before the next memory access.

\subsection{Runtime Reconfiguration}

\rvm{} is capable of doing user-level simulation, supervisor-level simulation and machine-level simulation. For user-level simulation, Linux syscalls are emulated, and for supervisor-level, \gls{SBI} calls are emulated.

In many cases, we want to gather cache statistics with the behaviour of \gls{OS} taken into account, but we do not want to count the \gls{OS} booting and workload preparation steps before the region of interest, and do not want to pay for the performance overhead of detailed models for these portions. The design of \rvm{} takes this into account, and both pipeline and memory models can be switched dynamically in the runtime. The switching is controlled by writing a special \gls{CSR} in the vendor-specific \gls{CSR} range.

\rvm{} supports pipeline model switching by simply flushing the code cache for translated binary, and let the \gls{DBT} engine to use the new model's hooks for code generation. Moreover, since as mentioned previously in \Cref{sec:overview}, each core has its own code cache for \gls{DBT}-ed code, we allow the pipeline models to be specified per core rather than at once.

The memory model is switched in the runtime by flushing the L0 data cache and the instruction cache. The cache line size is also a runtime-configurable property. For example, if both \gls{TLB} and cache are simulated, the cache line size can be set to 64 bytes. If only \gls{TLB} is simulated, the cache line size can be set to 4096 bytes, turning L0 data cache effectively into an L0 data \gls{TLB}.

If the memory model permits, \rvm{} can also switch between lockstep execution and parallel execution like other binary translators during the runtime. Parallel execution is enabled on the ``atomic'' memory model. When paired with the ``atomic'' pipeline model this behaves functionally equivalent to QEMU and gem5's atomic model which permits fast-forwarding of aforementioned booting and preparation steps.

\section{Evaluation}

As described in \Cref{sec:impl}, \rvm{} offers a range of pipeline models and memory models to select from, and allows switching between them mid-simulation. Each model shows different trade-offs. The list of pre-implemented pipeline and memory models can be found in \Cref{tab:pipe} and \Cref{tab:mem}.

\begin{table}[ht]
\centering
\begin{tabular}{|l|l|}
\hline
Name    & Description                               \\ \hline
Atomic  & Cycle count not tracked                   \\ \hline
Simple  & Each non-memory instruction takes one cycle           \\ \hline
InOrder & Models a simple 5-stage in-order scalar pipeline \\ \hline
\end{tabular}
\vspace{1\baselineskip}
\caption{List of pre-implemented pipeline models}
\vspace{-2\baselineskip}
\label{tab:pipe}
\end{table}

\begin{table}[ht]
\begin{tabularx}{\linewidth}{|l|X|}
\hline
Name   & Description                                                                                      \\ \hline
Atomic & Memory accesses not tracked                                                                      \\ \hline
TLB    & \gls{TLB} hit rate collected; cache not simulated                                                \\ \hline
Cache  & Cache hit rate collected; \gls{TLB} and cache coherency not modelled; parallel execution allowed \\ \hline
MESI   & A directory-based MESI cache coherency protocol with a shared L2. Lockstep execution required. \\ \hline
\end{tabularx}
\vspace{1\baselineskip}
\caption{List of pre-implemented memory models}
\vspace{-2\baselineskip}
\label{tab:mem}
\end{table}

\subsection{Accuracy and Validation}

For pipeline models, we validated the accuracy of the in-order model against an actual RTL implementation of a RISC-V core using CoreMark~\cite{coremark}. CoreMark is particularly helpful for this validation, as CoreMark's working set is small enough to fit into caches and therefore the memory system of the RTL implementation would not affect the benchmark result. In our run, the RTL implementation reports 2.10 CoreMark/MHz where the in-order model, when paired with the atomic memory model, reports 2.09 CoreMark/MHz. The difference is less than 1\%. The ``simple'' model is simply validated by checking that all cores have their MCYCLE and MINSTRET \gls{CSR} equal.

For memory models, we used a few micro-benchmarks to cover the use case for each model. For TLB and cache simulation, we used a single-core micro-benchmark that is similar to the MemLat tool from the 7-zip LZMA benchmark \cite{7zip}. For the MESI cache-coherency model, we used a micro-benchmark to simulate a scenario where two cores are heavily contending over a shared spin-lock. The memory model under test is used together with the validated in-order pipeline model, and we compare the number of cycles taken to execute a benchmark in \rvm{} and in RTL simulation. The error is around for the 10\% for the cache coherency model and lower for non-coherent models. Though not as accurate as the pipeline model, we believe at this accuracy the simulation can provide representative-enough metrics for exploring design decisions.

\subsection{Performance}

\begin{figure}[ht]
    \includegraphics[width=\linewidth]{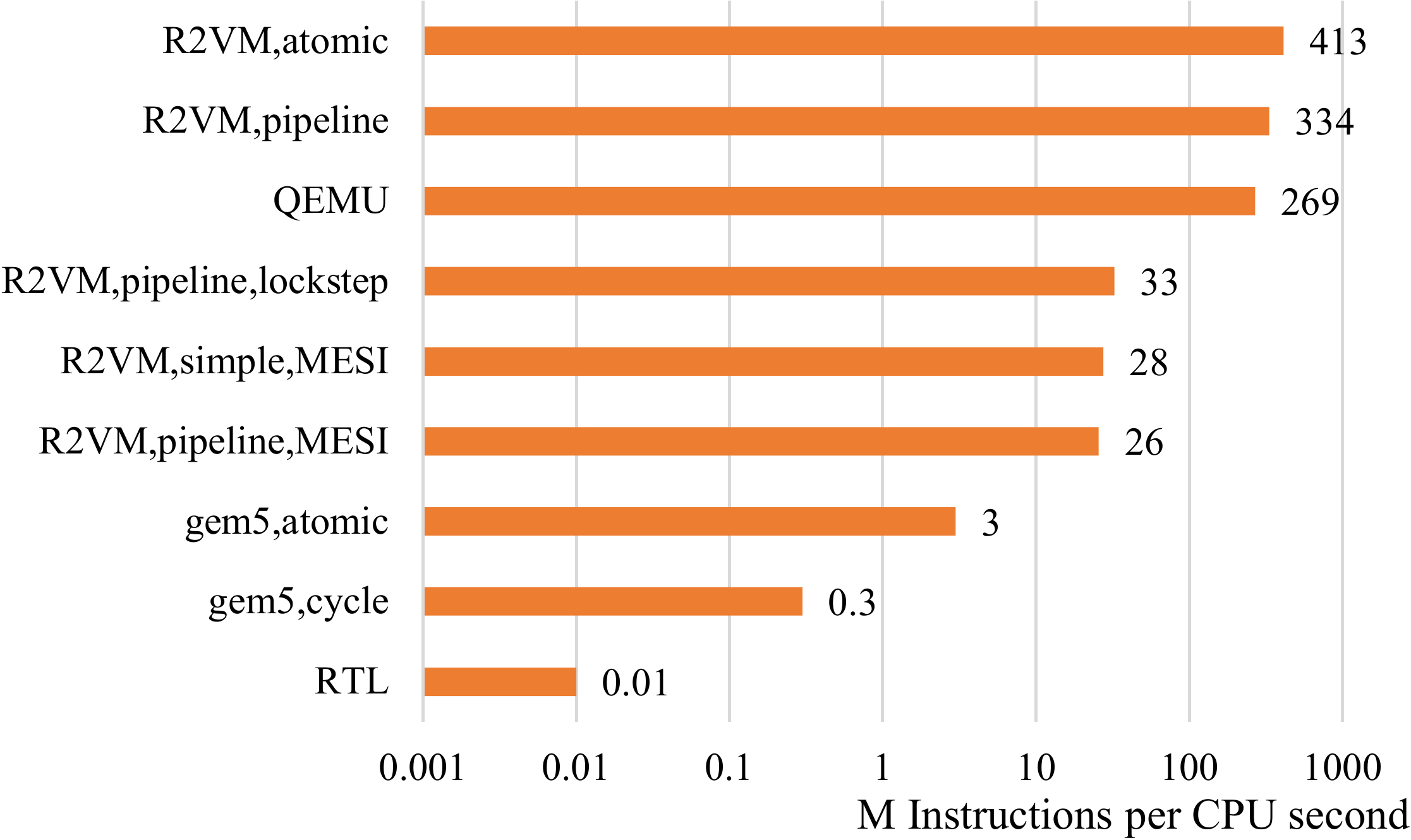}
    \caption{Performance comparison between models and other simulators}
    \label[figure]{fig:perf}
\end{figure}

We evaluated the performance of \rvm{} against QEMU using the deduplication workload from PARSEC \cite{bienia2008parsec} on 4 cores to test the integer performance of the simulator (as both \rvm{} and QEMU interprets floating-point operations). The kIPS numbers of the gem5 simulator are from Saidi et al.'s presentation \cite{saidi2012gem5}.

As shown in \Cref{fig:perf}, the techniques we use lead to superb performance. When caches are not simulated and therefore cores can run in parallel threads, \rvm{} runs at $>$300 MIPS per core, even outperforming QEMU. Lockstep execution brings down performance by 10x to $\sim$30 MIPS (for 4 simulated cores in a single-threaded), but this is still significantly faster than gem5. 

Thanks to our pipeline model design which moves most simulation to \gls{DBT} compilation time rather than runtime, and to our memory model design which offloads most memory accesses by using L0 caches, simulating pipelines and cache coherency protocols did not add a significant overhead themselves, compared to the overhead of lockstep execution.

\section{Conclusion}

We have introduced \rvm{}, a multi-purpose binary translating simulator that is able to simulate multi-core RISC-V systems at the cycle-level at high-speed. This is done by leveraging the use of fibers to support fast lockstep execution. Overall, optimisations made \rvm{} possible to achieve functional simulation performance that exceeds that of QEMU and cycle-level simulation nearly 100 times faster than gem5.

%
\bibliographystyle{ACM-Reference-Format}
\bibliography{paper}

\end{document}